\documentclass[pra,aps,twocolumn,superscriptaddress,amsmath,amssymb,floatfix]{revtex4}
\usepackage{graphicx}

\newcommand{\be}{\begin{equation}}
\newcommand{\ee}{\end{equation}}
\newcommand{\xx}{{\mathbf x}}

\newcommand{\rr}{{\mathbf r}}
\newcommand{\kk}{{\mathbf k}}

\newcommand{\eq}[1]{(\ref{#1})}
\newcommand{\mume}{\mu \rm m}

\begin{document}

\title{Coexisting Non-Equilibrium Condensates with Long-Range Spatial Coherence in Semiconductor Microcavities}

\author{D.~N.~Krizhanovskii}
 \affiliation{Department of Physics \& Astronomy, University of Sheffield, Sheffield S3 7RH, United Kingdom}

\author{~K.~G.~Lagoudakis}
 \affiliation{Ecole Polytechnique F\'ed\'erale de Lausanne (EPFL), Station 3, CH-1015 Lausanne, Switzerland}

\author{~M.~Wouters}
 \affiliation{Ecole Polytechnique F\'ed\'erale de Lausanne (EPFL), Station 3, CH-1015 Lausanne, Switzerland}

\author{~B.~Pietka}
 \affiliation{Ecole Polytechnique F\'ed\'erale de Lausanne (EPFL), Station 3, CH-1015 Lausanne, Switzerland}

\author{R.~A.~Bradley}
 \affiliation{Department of Physics \& Astronomy, University of Sheffield, Sheffield S3 7RH, United Kingdom}

\author{K.~Guda}
 \affiliation{Department of Physics \& Astronomy, University of Sheffield, Sheffield S3 7RH, United Kingdom}

\author{D.~M.~Whittaker}
 \affiliation{Department of Physics \& Astronomy, University of Sheffield, Sheffield S3 7RH, United Kingdom}

\author{M.~S.~Skolnick}
 \affiliation{Department of Physics \& Astronomy, University of Sheffield, Sheffield S3 7RH, United Kingdom}

\author{B.~Deveaud-Pl\'{e}dran}
 \affiliation{Ecole Polytechnique F\'ed\'erale de Lausanne (EPFL), Station 3, CH-1015 Lausanne, Switzerland}

\author{M.~Richard}
 \affiliation{Institut N\'eel, CNRS and Universit\'e J. Fourier, 38042 Grenoble, France}

\author{R.~Andr\'e}
 \affiliation{Institut N\'eel, CNRS and Universit\'e J. Fourier, 38042 Grenoble, France}

\author{Le Si Dang}
 \affiliation{Institut N\'eel, CNRS and Universit\'e J. Fourier, 38042 Grenoble, France}

\date{January 29, 2009}

\begin{abstract}
Real and momentum space spectrally resolved images of microcavity polariton emission in the regime of condensation are investigated under non resonant excitation using a laser source with reduced intensity fluctuations on the timescale of the exciton lifetime. We observe that the polariton emission consists of many macroscopically occupied modes. Lower energy modes are strongly localized by the photonic potential disorder on a scale of few microns. Higher energy modes have finite k-vectors and are delocalized over 10-15 $\mu$m. All the modes exhibit long range spatial coherence comparable to their size.
We provide a theoretical model describing the behavior of the system with the results of the simulations in good agreement with the experimental observations. We show that the multimode emission of the polariton condensate is a result of its nonequilibrium character, the interaction with the local photonic potential and the reduced intensity fluctuations of the excitation laser.
\end{abstract}


\maketitle


\section{Introduction}

There is significant contemporary interest in the study of strongly coupled semiconductor microcavities, where mixed exciton-photon quasi-particles, two-dimensional (2D) polaritons with very small effective mass can be created~\cite{Weisbuch1995,Kavokin2007,Deveaud2007}. As a result, high density macroscopically occupied polariton states can be achieved at high temperatures and relatively small optical excitation densities. Most notably, the formation of macroscopically occupied states arising from polariton Bose-Einstein Condensation (BEC), has been recently observed in CdTe ~\cite{Kasprzak2006}, GaAs~\cite{Balili2007, Krizh2007}, and GaN ~\cite{Baumberg2007,christman08}  microcavities under conditions of non-resonant excitation. Polariton condensates exhibit the characteristic properties of BEC, such as long range spatial coherence ~\cite{Kasprzak2006}, suppresion of the photon bunching effect at threshold ~\cite{Kasprzak2008}, long coherence time for both the first (g1) and second (g2) order correlation functions ~\cite{Love2008} and the quantization of vortices ~\cite{Lagoud2008}. Although there is evidence for equilibrated polariton distributions ~\cite{Kasprzak2006, Balili2007,Deng2006},  and macroscopic occupation of the k=0 ground state, polariton condensates are far from thermodynamical equilibrium: due to the rather small polariton lifetimes these states originate from a dynamical balance of pumping and losses in the system. There are several approaches based on both microscopic calculations and generalization of the Gross-Pitaevskii equation, which describe the consequences of the non-equilibrium aspect of the interacting polariton BEC system~\cite{Wouters2007, Szym2006, Keeling, Savona}.

The imperfections of semiconductor microcavity structures result in a disordered photonic potential landscape, which plays an important role in polariton propagation and especially, formation of 2D macroscopically occupied polariton states. The influence of photonic disorder was first discussed for the case of the optical parametric oscillator ~\cite{Stev2000}, where macroscopically occupied signal and idler states are formed due to direct scattering of the resonantly excited pump polaritons. It was shown that the photonic disorder strongly affects the real space distribution of the signal~\cite{Sanvitto2006}, which consists of spatially localized modes emitting at different energies~\cite{Krizh2006}. The non-equilibrium polariton BEC in CdTe microcavities excited non-resonantly was also observed to consist of several localized maxima~\cite{Kasprzak2006}, which arise from polariton trapping by the potential disorder. In Ref.~\cite{Baas2008}, it was demonstrated that synchronization between condensates localized in neighboring minima is possible. In addition, it was shown that polariton condensates subject to a disorder potential may spontaneously exhibit quantized vortices due to interplay between the disorder potential and the flow and decay of the condensate~\cite{Lagoud2008}.

Recent experiments revealing very long coherence times of the polariton BEC using CW diode laser excitation in CdTe structures showed that microcavity polariton emission consists of multiple narrow modes strongly overlapping in real space~\cite{Love2008}. Such an observation became possible, since the diode laser has reduced intensity noise on a timescale of 100-1000 ps and results in reduced fluctuations of the number of excitons and polaritons, which otherwise lead to marked broadening of BEC spectra and prevent the observation of the coexisting condensates~\cite{Love2008}.

\begin{figure}[!hb]
\begin{center}
\includegraphics[width=0.9\columnwidth,angle=0,clip]{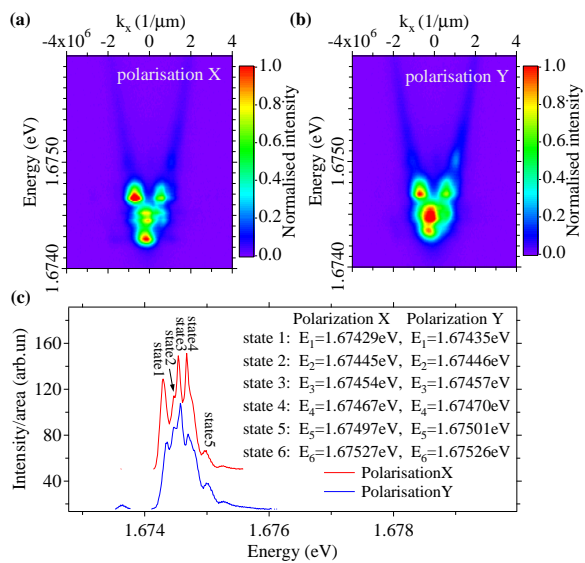} \end{center}
\caption{(a),(b)Dispersions of the lower polariton branch for two perpendicular polarizations X and Y.(c)The real space intensity of the luminescence averaged over mode size.}
\label{fig:Disp}
\end{figure}
In the present work, we present spectrally resolved real and momentum space images of the multimode polariton condensate and probe the properties of the spectrally resolved long range spatial coherence of each mode.  We observe that above threshold condensation occurs into several polariton levels. This is a result of the non-equilibrium aspects of the system, which prevents thermodynamical equilibrium being reached for the states near the bottom of the lower polariton (LP) branch. The low energy condensate states are found to be strongly localized within a deep potential minimum, whereas the higher energy states are spatially extended over few potential minima. In momentum space, these delocalised states typically consist of a few bright maxima distributed on a ring. Similar to the phenomenon of coherent back scattering of resonantly excited disordered systems, a peak at some $\kk$ is often accompanied by a weaker satellite at $-\kk$.

\begin{figure}[!hb]
\begin{center}
\includegraphics[width=0.9\columnwidth,angle=0,clip]{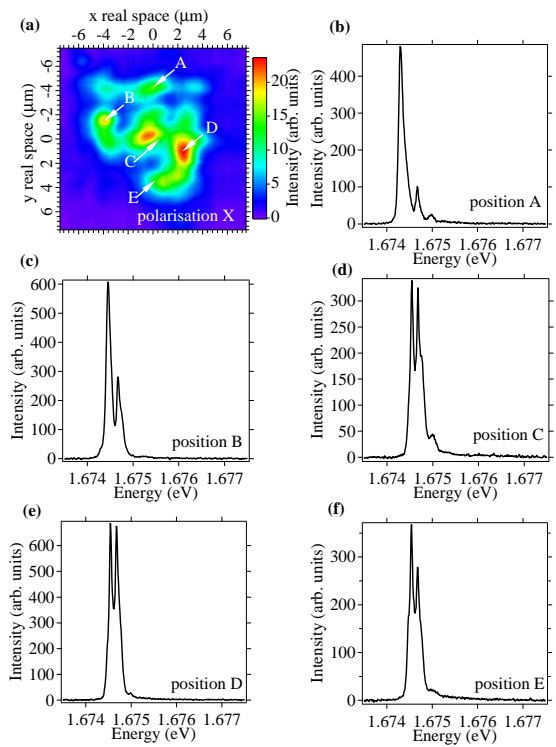} \end{center}
\caption{(a)real space spectrally averaged luminescence image for X polarization. (b)-(f)local spectra obtained from small regions in real space at positions shown by the arrows in panel (a).}
\label{fig:SpatialPeaks}
\end{figure}

After the discussion of the experimental observations in Sec.~\ref{sec:exp}, we present in Sec. \ref{sec:theor} a theoretical analysis of a multiple frequency condensate based on a mean field model for polariton condensation~\cite{Wouters2007}, taking into account the finite life time and the disorder acting on the polaritons.
Very  good qualitative agreement between theory and experiment is observed.
Conclusions are drawn in Sec.~\ref{sec:concl}

 \section{Experiment \label{sec:exp}}

The sample employed here is the one used in Ref.~\cite{Kasprzak2006} where a BEC with extended spatial coherence was reported. The sample was cooled to $10^{\circ}K$. To avoid heating of the sample, quasi-CW non-resonant excitation was employed by means of two CW diode lasers at 685 nm using a mechanical optical chopper with a frequency of $300$ Hz. The excitation conditions are similar to those employed in Ref~\cite{Love2008}, where multimode emission of polariton BEC was observed.
The size of the excitation spot was about 20 $\mu$m. High resolution sub micron imaging was achieved by a high numerical aperture microscope objective (N.A=0.5). Spectrally and spatially resolved images were recorded using a double monochromator of about 30 $\mu$eV resolution and a CCD camera. Interferometric measurements were performed by means of a modified Michelson interferometer in a mirror-retroreflector configuration with active stabilization~\cite{Kasprzak2006, Baas2008, Lagoud2008}.

\subsection{Spectrally resolved images in real and momentum space \label{sec:images}}
\begin{figure}[!ht]
\begin{center}
\includegraphics[width=0.9\columnwidth,angle=0,clip]{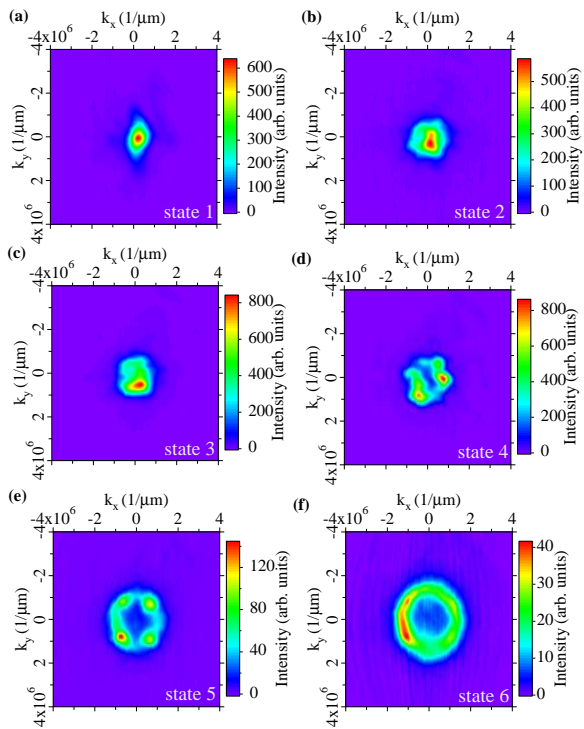} \end{center}
\caption{(a)-(f): state 1-state 6 momentum space imaging at the energy of each mode.}
\label{fig:KpolX}
\end{figure}

\begin{figure}[!ht]
\begin{center}
\includegraphics[width=0.9\columnwidth,angle=0,clip]{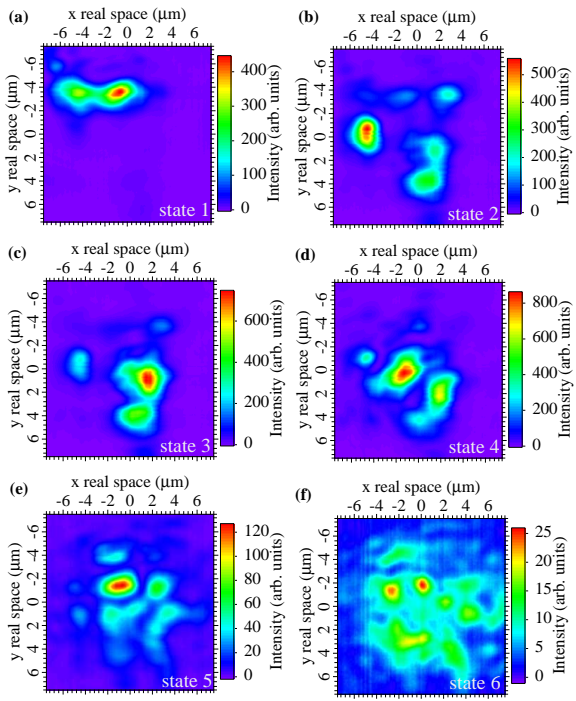} \end{center}
\caption{(a)-(f): state 1-state 6 real space imaging at the energy of each mode.}
\label{fig:RpolX}
\end{figure}

At low excitation powers below the condensation threshold  polariton emission is quite broad (FWHM$~1-1.5$ meV) with no resolved spectral features, as reported previously~\cite{Kasprzak2006,Love2008}.
By contrast, above condensation threshold strong polariton emission arises, which consists of multiple narrow lines~\cite{Love2008}.
Fig.\ref{fig:Disp}(a),(b) show the images (dispersions) of polariton condensate emission above the condensation threshold in energy-momentum (E-$k_{x}$) space taken at momentum $k_{y}=0$ for $P=1.4$ mW $\approx2 P_{th}$ excitation power, where $P_{th}$ is the condensation threshold. The spectra are recorded for two perpendicular linear polarizations X and Y which are respectively parallel and perpendicular to the crystallographic axis $\langle1,\overline{1},0\rangle$ of the sample. The polarization degree is of the order of $80$ percent with the strongest emission in X-polarisation, which is consistent with our previous observation~\cite{Kasprzak2006}.
Fig.\ref{fig:Disp}(c) shows the corresponding spectrum obtained by collecting light from the 20 $\mu$m excitation area. We are able to distinguish 6 narrow peaks (states 1-6) in the polariton emission of each polarization. The energies of these peaks are separated by  $100-250\mu$eV. A splitting of $\approx25\mu$eV between the X and Y polarized narrow lines is observed for each of the modes. Such a splitting also exists for the LP mode below threshold and is probably due to intrinsic anisotropy in the microcavity mirrors, leading to a cavity birefringence.
Only the first 4 lower energy peaks (states 1-4) are observed in Fig.\ref{fig:Disp}(a),(b) in (E-$k_{x}$) space, since the higher energy peaks (states 5-6) have maxima of intensity at $k_{y}>0$. However, these peaks are clearly revealed in 2D images in momentum space as shown in Fig.\ref{fig:KpolX}.

In order to reveal further information on the mode structure without complication of spatial averaging as in Fig.\ref{fig:Disp}(c), spectra were recorded with spatial selection from specific regions of $\sim1\mu$m$^2$ size across the excitation spot. The real space spectrally averaged image of the condensate is shown in Fig.\ref{fig:SpatialPeaks}(a).  Fig.\ref{fig:SpatialPeaks}(b)-(f) shows X-polarized condensed polariton emission spectra obtained for detection at each of the points labeled A-E in Fig.\ref{fig:SpatialPeaks}(a). Different peaks dominate depending on the detection area across the spot. However,  as seen in Fig.\ref{fig:SpatialPeaks}(b)-(f) the emission from small regions always consists of about $3$ peaks, which suggests that the modes of the condensate strongly overlap in space.

In both polarizations the peaks are narrow: their linewidths (FWHM) lie in the range from $150$ $\mu$eV down to $50$ $\mu$eV (resolution limited) for the most intense states 3 and 4 ~\cite{CommentOnCoherence}. Such an observation is consistent with our previous studies, where long coherence times up to $200$ ps (FWHM$=10$ $\mu$eV) were reported for a single mode of microcavity polariton emission above condensation threshold. Strong spectral narrowing and superlinear increase of the intensity of each single mode is observed with increasing excitation power~\cite{Love2008}, which indicates the build-up of coexisting macroscopically occupied states (polariton condensates) and condensation into both non-degenerate X and Y polarized states.

The number of condensates observed varies from $1$ up to $6-7$, and this, as well as the energy separation between the condensed modes, depends strongly on the excitation position across the sample. As we show in this manuscript the multimode emission above threshold arises from interaction of polaritons with the local photonic disorder potential.

\begin{figure}[!t]
\begin{center}
\includegraphics[width=0.9\columnwidth,angle=0,clip]{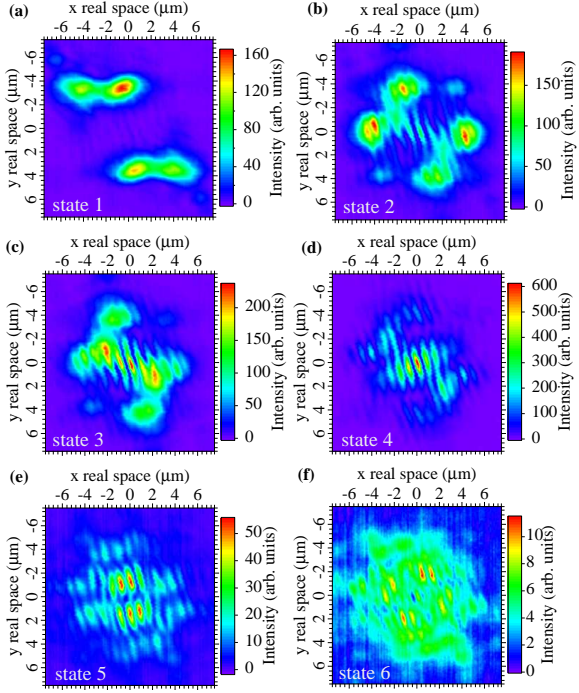} \end{center}
\caption{(a)-(f)real space imaging of the interference pattern of modes state 1-state 6.}
\label{fig:INTpolX}
\end{figure}

 \begin{figure}[!t]
\begin{center}
\includegraphics[width=0.9\columnwidth,angle=0,clip]{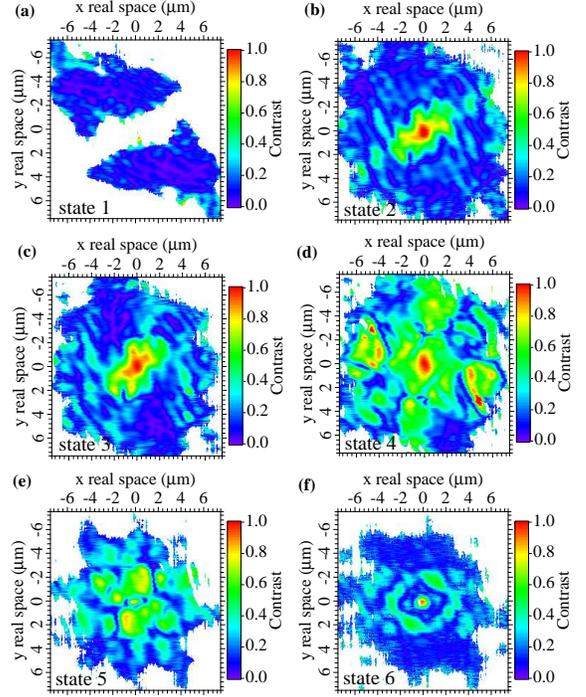} \end{center}
\caption{(a)-(f) contrast of the interference pattern of each mode. In the blanc regions, the measured intensity is very low giving a non defined contrast.}
\label{fig:CNTpolX}
\end{figure}

\begin{figure}[!hb]
\begin{center}
\includegraphics[width=0.9\columnwidth,angle=0,clip]{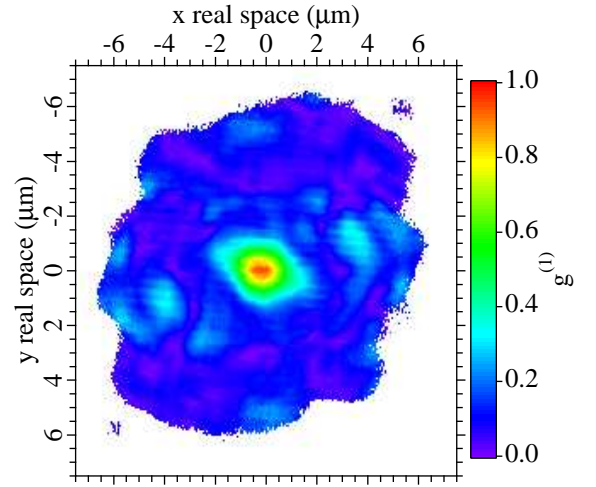} \end{center}
\caption{(a),(b) real space correlation pattern. The blanc regions are the regions where there is no intensity and thus the correlations lead to a high amplitude noise }
\label{fig:g1XY}
\end{figure}

To further reveal the origin of formation of the spectrally narrow coexisting condensates we recorded spectrally and spatially resolved 2D images in real and momentum space. The images were obtained above threshold for each of the narrow peaks in Fig.\ref{fig:Disp}(c). Although the Y-polarized modes are spectrally distinct from the corresponding X-polarized modes, they are observed to have an almost identical pattern in both real and momentum space. Such a similarity arises from the fact that the photonic potential, which determines the new polariton condensed states, is given by the spatial fluctuations of optical thickness of the cavity due to defects in the distributed Bragg reflectors and thus is identical for both X and Y polarized polariton modes. Therefore, here we present the study of only one X polarization.

Fig.\ref{fig:KpolX} (a-f) shows spectrally resolved 2D images recorded in momentum space for X-polarized modes, each panel corresponding to the energy of each mode for the same excitation conditions and spot size as in Fig.\ref{fig:Disp}(c). It is seen that the lower energy states are localized around $k=0$. On the contrary, higher energy condensed states have well defined k-vectors different from zero. State 1, which is the  ground state mode emitting at $1.67429$ eV, is localized around $k=0$ within $\pm 0.6$ $\mu$m$^{-1}$. The second and third modes (state 2, state 3) emitting at energies  $1.67445$eV and $1.67455$eV, respectively, show increasing spreading in k space. The fourth and fifth modes show multiple lobe structure (state 4, state 5) and the fifth mode has four well defined symmetrically positioned lobes. The lobes have wavevectors $(k_{x},k_{y})\approx(-1,-1)\mu m ^{-1}$, $(k_{x},k_{y})\approx(1,1)\mu m ^{-1}$ and $(k_{x},k_{y})\approx(1,-1)\mu m ^{-1}$,$(k_{x},k_{y})\approx(-1,1)\mu m ^{-1}$. The highest energy mode (state 6) is the weakest and is found on a ring in k-space with most of the population being at the negative $k_{x}$ side.
Simultaneous appearance of condensate density at $\pm \kk$ was observed in numerous other k-space images that we recorded (not shown) and also in theoretical simulations (see below). The satellites at $-\kk$ may be related to coherent backscattering of the polaritons that condense at $+\kk$ due to its interaction with the local photonic potential.


Fig.\ref{fig:RpolX}  shows spectrally resolved images in real space recorded for each of the narrow modes for the dominant X polarization. The lowest energy mode (state 1) is strongly localized in real space within an area of $\approx6\times2$ $\mu$m$^2$ , whereas the wavefunctions of the higher energy polariton states are observed to consist of several maxima separated by $\approx2-8$ $\mu$m and are extended over a lengthscale $10-15$ $\mu$m . Moreover all the modes show strong spatial overlap, indicating coexistence of the condensed phases.

As it is shown below there is a long range spatial coherence established for each of the spectrally resolved condensate modes over distances comparable to their size. Therefore, the images of the modes in k-space are expected to be Fourier transforms of the corresponding images in real space ~\cite{FT}. In other words, the k-space patterns arise from  interference between emission of the same energy originating from different regions in real space.  The fact that the higher energy modes (states 4-6) have maxima of intensity at k-vectors different from zero suggests that there is a large phase difference between polariton fields, which originate from different areas (maxima) in the corresponding real space images. The k-vectors of the higher energy condensed modes are, thus, expected to correspond to a wavelength of the order of the separation between the maxima in agreement to experimental observations. This is further discussed in the theory part of this manuscript.

\subsection{Spatial coherence of spectrally resolved condensate modes \label{sec:coherence}}

A defining feature of a Bose-Einstein condensate is long range spatial coherence. As in previous experiments on the same sample, the  correlations of the emitted luminescence in real space were measured by means of an actively stabilized  Michelson interferometer in a mirror-retroreflector configuration~\cite{Kasprzak2006, Baas2008, Lagoud2008}. In order to probe the coherence properties of every mode separately, we  measured the correlations of \textit{each individual mode} in real space. The interferogram at the output of the interferometer was directed to the entrance slits of the monochromator and then the real space interferogram was spectrally resolved. The interferogram that we acquired in real space allowed us to gain access to the correlations for each state. Fig.\ref{fig:INTpolX} shows the reconstructed interference pattern for each of the modes. The extracted contrast of each state in real space reveals the coherent content of each of the modes. Fig.\ref{fig:CNTpolX} shows the contrast for polarization X, calculated from the interference pattern. The existence of highly correlated polariton populations in each of the modes, indicates formation of coherence over a length comparable to the mode size ($5-15$ $\mu$m). The lack of correlations for the ground state mode (state 1) is not due to the lack of a coherent polariton population but rather due to the positioning and strong localization of this mode in real space, which results in no overlap between the mode and its retroreflected image.

 Fig.\ref{fig:g1XY}  shows energy averaged correlations in real space. The correlations are observed to be very low and extend over only $2-3$ $\mu$m. Such an observation arises from the fact that the condensed modes have different momentum distributions and, thus, the momentum distribution of the total polariton emission is quite broad. Therefore, spatial interference patterns arising from each of the condensed modes are different and may be expected to cancel out in the case of energy-averaged detection. However, we note that it is possible to obtain long range spatial coherence for energy averaged emission of polariton condensates~\cite{Kasprzak2006}, if one selects a position on the sample for which a single polariton condensed mode with a relatively narrow  momentum distribution dominates above threshold. In previous experiments ~\cite{Kasprzak2006}, attention was always paid to choose a position on the sample with the smallest possible disorder and long spatial coherence. Here we present the results for a generic position on the sample with larger amplitude disorder.


As the final point of this section, we point out that multimode structure is a feature that was not revealed when the condensate was created on the same sample with a multimode Ti:Sapph laser~\cite{Kasprzak2006}, because in that case the condensate linewidth was larger than the mode spacing. The relevant difference between the Ti:Sapph and the diode laser is the time scale of intensity fluctuations. For the Ti:Sapph laser, the mode spacing is of the order of 250 MHz, resulting in intensity fluctuations on ns timescale. The intensity fluctuations of the diode laser are much faster (ps timescale) due to their much larger mode spacing ($\approx$ 25 GHz). The importance of the intensity fluctuation time scale can be understood in the following way. Slow laser intensity fluctuations are followed by the exciton reservoir (relaxation time of the order of ns), whereas fast fluctuations are averaged out. A fluctuating reservoir density results in a temporal variation of the blue shift and hence a linewidth broadening. This explains why the condensate linewidth under Ti:Sapph excitation (hundreds of $\mu$eV) is much larger than the one under diode laser excitation (tens of $\mu$eV)~\cite{Love2008}.

\section{Theoretical description \label{sec:theor}}

The simultaneous occurrence of macroscopically occupied polariton states at different frequencies is only possible out of equilibrium: the condensate frequency corresponds to the chemical potential~\cite{bec-book}, which for an equilibrium system is unique and spatially homogeneous. Considering the short life time of polaritons (a few ps), the non-equilibrium character is not surprising: e.g. the thermal velocity does not allow a polariton to travel more than a few microns within its life time, a distance that is not sufficient to sample the whole excitation area. Polariton condensates share this non-equilibrium character with conventional lasers.
 \begin{figure}[!t]
\begin{center}
\includegraphics[width=1.\columnwidth,angle=0,clip]{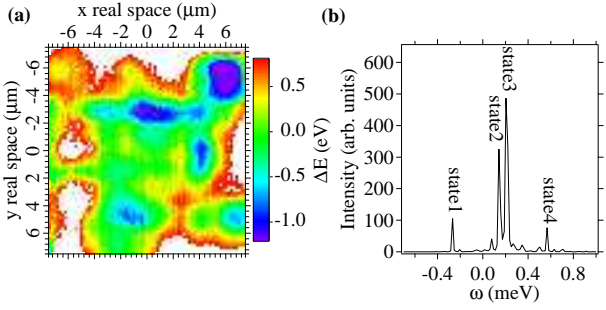} 
\end{center}
\caption{(a)Disorder potential landscape used in the theoretical simulations and (b)energy spectrum from a simulation with the mean field model \eq{eq:GP} and \eq{eq:rate}. Multiple condensates are found in the simulation. The zero of the energy is the same as in (a).
Values of parameters used in the simulations: $\hbar g=0.015\, \textrm{meV} \,\mume^2$, $\hbar\gamma_c=0.5\,\textrm{meV}$, $\hbar\gamma_R=10\,\textrm{meV}$, $\hbar R[n_R]=( \textrm{meV} \mume^2) \times n_R$  $\hbar\,g_R=0$, $\mathcal G = 0$ and $P/P_{th}=2$.}
\label{fig:VD_CONDW}
\end{figure}

\begin{figure}[!bl]
\begin{center}
\includegraphics[width=1.\columnwidth,angle=0,clip]{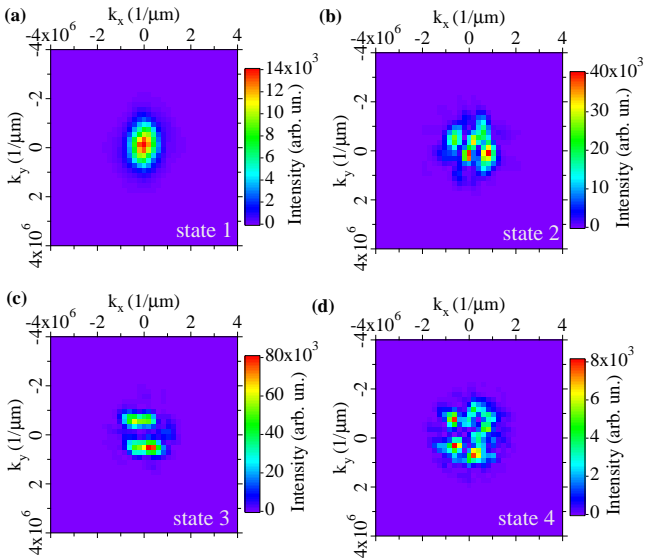}
\end{center}
\caption{Momentum space distribution of the polariton field at the corresponding peak frequencies in Fig. \ref{fig:VD_CONDW}(b).}
\label{fig:condk}
\end{figure}
The crucial difference between a polariton condensate and an ordinary laser lies rather on the microscopic than on the macroscopic level: a polariton condensate does not require population inversion, because the statistics of collisions ensures the relaxation of hot excitons, avoiding the need for population inversion. On the macroscopic level on the other hand, there are no fundamental differences and there is a strong analogy between polariton condensates in random potentials and lasing in random materials~\cite{random-laser}.

\begin{figure}[!br]
\begin{center}
\includegraphics[width=1.\columnwidth,angle=0,clip]{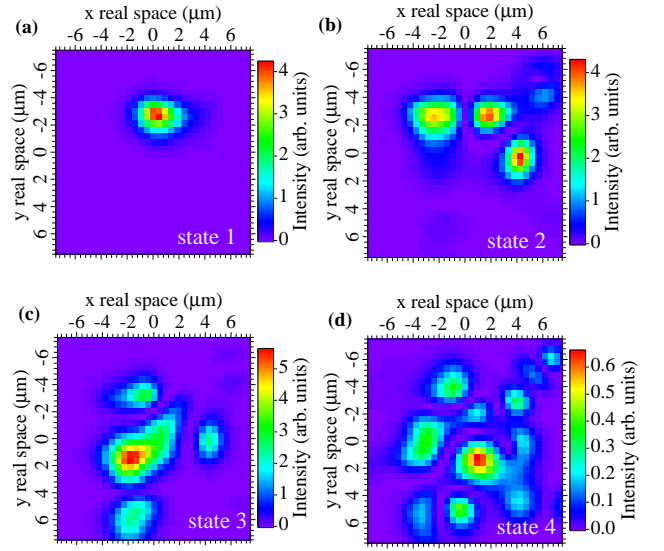} 
\end{center}
\caption{Real space distribution of the polariton field at the corresponding peak frequencies in Fig. \ref{fig:VD_CONDW}(b).}
\label{fig:condx}
\end{figure}

Given the good spatial coherence of each individual condensate mode (see Fig.\ref{fig:CNTpolX}), it is justified to neglect in our theoretical description the fluctuations and to study the formation of multiple condensates within a mean field model \cite{Wouters2007,Keeling}. We will adopt here the equations introduced in Ref. \cite{Wouters2007}, that were used to explain the effects of a finite size pump spot~\cite{Wouters2008} and the appearance of quantized vortices~\cite{Lagoud2008} in polariton condensates.
It consists of a Gross-Pitaevskii equation including losses and gain for the macroscopically occupied polariton field $\psi(\xx)$:
\begin{multline}
i\hbar \frac{\partial \psi(\rr)}{\partial t}=\Big\{E_0 -\frac{\hbar^2}{2m} \nabla_\rr^{2}+
\frac{i\hbar}{2}\big[R[n_R(\rr)]-\gamma_c\big] \\
+V_{ext}(\rr)+\hbar g\,|\psi(\rr)|^{2}+V_R(\rr) \Big\}
\psi(\rr), \label{eq:GP}
\end{multline}
where $E_0=\hbar\omega_0$ and $m$ are respectively the minimum energy and the
effective mass of the lower polariton branch and  $g>0$ quantifies the
strength of repulsive binary interactions between condensate
polaritons. Exciton and cavity disorder can be included
as an external potential term $V_{ext}(\rr)$. At the simplest
level, the corresponding gain rate $R[n_R]$ can be described by a
monotonically growing function of the local density $n_R(\rr)$ of
reservoir excitons in the so-called bottleneck
region~\cite{porras}. At the same time, the reservoir produces a
mean-field repulsive potential $V_R(\rr)$ that can be approximated by
the linear expression $V_R(\rr)\simeq \hbar g_R\,n_R(\rr)+\hbar
{\mathcal G}\,P(\rr)$, where $P(\rr)$ is the (spatially dependent)
pumping rate and $g_R,{\mathcal G}>0$  are phenomenological coefficients
The GPE equation \eq{eq:GP} for the condensate has then to be coupled to
a rate equation for $n_R(\rr)$~:
\begin{equation}
\dot{n}_R(\rr)=P(\rr)-\gamma_R\,n_R(\rr)-R[n_{R}(\rr)]\,|\psi(\rr)|^{2}, \label{eq:rate}
\end{equation}
where $P$ describes the filling of the reservoir by the non-resonant excitation.

Let us recapitulate the solutions of the equations \eq{eq:GP} and \eq{eq:rate} in the spatially uniform case $V_{ext}=0$. For low excitation density $P$, a stable steady state is given by $\psi=0$ and $n_R=P/\gamma_R$.
The exciton density reaches the threshold for polariton condensation when the gain from stimulated scattering from the exciton reservoir into the lower polariton branch equals the polariton loss rate: $R(n_R)=\gamma_c$. For pump powers above the threshold $P>P_{th}=\gamma_R n_R^{th}$, the solution without a condensate $\psi=0$ becomes dynamically unstable. The dynamically stable solution is then given by the condition that the gain is clamped to the losses $R[n_R]=\gamma_c$ and the condensate density is $|\psi|^2=(P-P_{th})/\gamma$. The condensate wave function oscillates at the frequency $\omega_c=g |\psi|^2+V_R$.

In the presence of an external potential, this simple picture breaks down. In particular, it is no longer guaranteed that only a single frequency appears above the condensation threshold. In the context of polariton condensation multiple frequency solutions of the related complex Ginzburg Landau equation have been found for certain types of regular external potentials~\cite{eastham2008}.

First of all, the photonic potential leads to the localization of the lower energy polariton modes with small k-vectors on a scale of few microns. However, there is a continuum of polariton states with higher k-vectors, which have kinetic energies above the amplitude of the photonic potential disorder. The losses can be overcome by the gain for many modes with different energies, which may result in simultaneous condensation into both localized and delocalized states.

The disorder potential that we used for the theoretical calculations is shown in Fig. \ref{fig:VD_CONDW}(a). The shape of the disorder potential is determined from the spatial distribution of the low pump intensity photoluminescence. The spatial variation of the lowest emission energy follows the disorder potential, albeit not perfectly because of the zero point kinetic energy that smooths out the variations of the potential energy. To compensate for the effect of the kinetic energy, we have multiplied the variations in the experimentally observed energy landscape by a heuristic factor of 3.5 so as to reproduce approximately the separation between the condensate frequencies.

The solution of our model equations \eq{eq:GP} and \eq{eq:rate} is shown in Figs. \ref{fig:VD_CONDW}(b)-\ref{fig:condk}. As in the experiment, multiple frequencies are present in the spectrum. The observation of multiple frequencies is consistent with the parameters used in the simulations: the fluctuations in the disorder potential are larger than the blue shift due to condensate-condensate interactions $g |\psi|^2$ (its maximal value in the simulation is 0.3 meV). The coupling between the various modes at different energies is therefore too weak to ensure synchronization~\cite{eastham2008}. The real and reciprocal space distributions at the peaks in Fig.~\ref{fig:VD_CONDW}(b) are shown in Figs. \ref{fig:condx} and \ref{fig:condk} respectively.

First of all, we want to draw the attention to the strong qualitative similarities between the theoretical simulations and the experimental measurements. In both cases, the lowest energy states are very localized in real space (extended over a few microns). In contrast, the states at higher energy are delocalized: they are extended over a much larger area, with a diameter of the order of 10 $\mume$.
\begin{figure}[!t]
\begin{center}
\includegraphics[width=1\columnwidth,angle=0,clip]{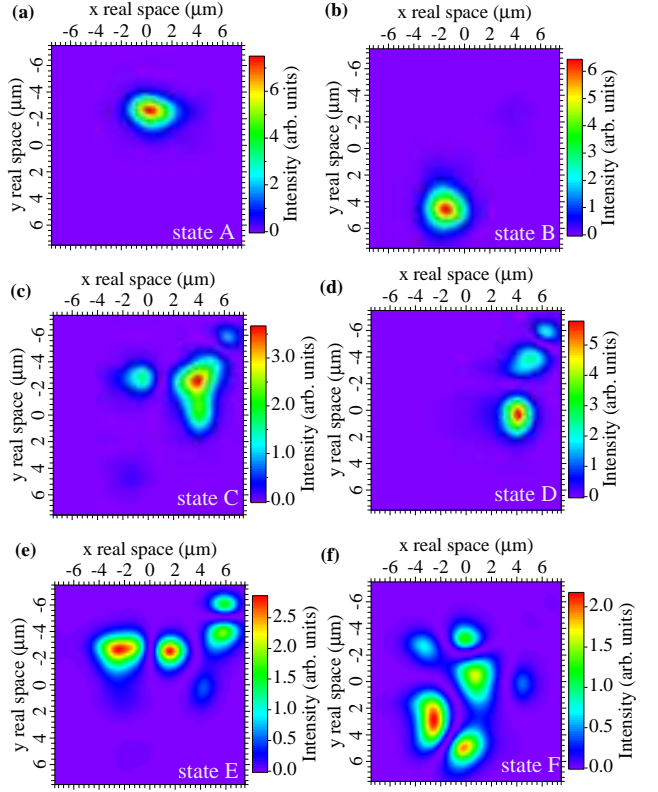} 
\end{center}
\caption{Real space density of the calculated lowest energy microcavity eigenstates that are localized within the excitation region. Modes A and B and C can be identified with the three lowest condensate modes.}
\label{fig:linearx}
\end{figure}
The qualitative correspondence between the experimental and theoretical reciprocal space images is also excellent. The low energy states form a disk around $\kk=0$, while the states at highest energy lie on a ring in momentum space. Note that all the modes except the lowest one do not exhibit inversion symmetry in $k$-space. This lack of inversion symmetry is also present in the experimental momentum distributions above threshold: we have observed many peaks in the momentum distribution which have a satellite at the opposite wave vector. This is likely to be related to the phenomenon of backscattering in disordered systems.

Time reversal of the linear Schr\"odinger equation requires that the eigenstates in the disorder potential are invariant under $\kk\rightarrow -\kk$. The condensate modes therefore do not coincide with the linear eigenstates of the microcavity.
\begin{figure}[!t]
\begin{center}
\includegraphics[width=1\columnwidth,angle=0,clip]{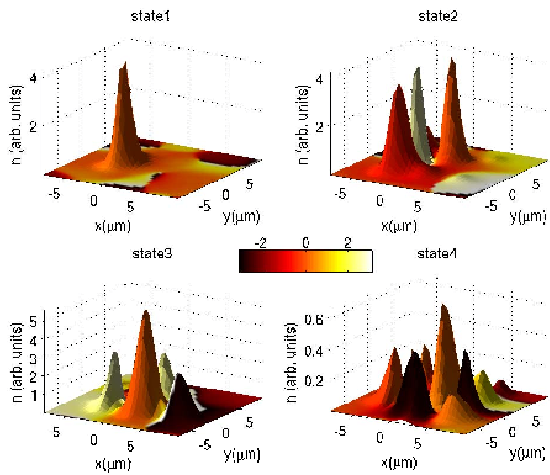}
\end{center}
\caption{Simultaneous representation of the density (hight of the surface) and phase (color coding, lighter is higher phase) of the polariton condensate modes. Note that the phase difference between neighboring maxima is typically large.}
\label{fig:flow}
\end{figure}
An identification of the condensate states with the linear eigenstates (see Fig.~\ref{fig:linearx}) can only be made for the lowest condensate state (it corresponds with the lowest eigenstate A).  The higher energy states, on the other hand cannot be identified with a linear eigenstate. Instead, the nonlinear dynamics of pumping, dissipation and polariton-polariton interactions defines the new modes into which the polaritons condense.

We now discuss what selects the energies and k-vectors of the delocalized modes at which polariton condensation is triggered. To obtain further insight into the nature of the condensate states, it is instructive to look simultaneously at the density and phase of the polariton modes. The height of the surface of Fig.\ref{fig:flow} represents the density, where the color represents the phase. Within one maximum of the polariton density, the phase is almost constant. The phase difference between  the peaks is by contrast large (of the order of $\pi$). These phase differences allow for currents between the different potential minima in order to lock to the same frequency~\cite{sync_theory}. These states are reminiscent of the excited $\pi$ states that were observed with polariton condensates in a periodic instead of a disordered potential~\cite{yamamoto_pi}. Josephson currents are associated with the large phase difference between the different condensate regions. They flow from high (light colored) to low (dark colored) phase. The large phase difference between the polaritons at the different potential minima explains why the radius of the momentum distributions corresponds to a wavelength of the order of the separation between the potential minima. Indeed, as seen from the theoretical and the experimental real space images of the condensed states (Fig.\ref{fig:RpolX} and Fig.\ref{fig:condx}, respectively) the typical separation $d$ between the intensity maxima in real space is about $3-4$ $\mu$m. Therefore,  taking into account the large phase difference the emission in the far-field due to interference is expected to have maxima at angles $\theta$ with respect to the normal of the samples such as $d\sin(\theta)=\lambda/2$, where $\lambda$ is the wavelength of condensate emission. These angles correspond to the values of k-vectors $\kk=\pi/d \sim1 \mu$m$^{-1}$ in agreement with the experimental and theoretical k-space images (Fig.\ref{fig:INTpolX} and Fig.\ref{fig:condk}, respectively).
Note that in the second state (state 2, Fig.\ref{fig:condx}(c)), the flow of polaritons is directed from
the outside to the center, thereby avoiding in plane losses of
polaritons out of this condensate mode. This idea of macroscopic
occupation of states that minimize the losses has been put forward in
the theory of random lasers ~\cite{random-laser}. However, in the present system, such a
a mechanism does not seem to be important: the flow in the third state (state 3, Fig.\ref{fig:condx}(c))
goes in the outward direction, yet the largest number of polaritons is
condensed in this state.

\section{Conclusions \label{sec:concl}}

We have presented a detailed experimental and theoretical analysis of a multimode polariton condensate. The real and reciprocal space densities of the different modes were recorded. We have observed that the lowest energy mode is typically localized within a single deep potential minimum, whereas the higher energy modes are delocalized over many potential minima. In reciprocal space, the lowest energy mode is centered around $k=0$; the higher energy modes tend to have a $k$-space density distributed on a ring with a radius that corresponds to a wavelength of the order of the distance between the potential minima.
Due to the breaking of time reversal, the condensate modes lack inversion symmetry in reciprocal space, although often a peak observed in the momentum distribution at given $\kk$ is accompanied by a weaker peak at $-\kk$, a phenomenon which may be explained by the occurrence of coherent backscattering in the disorder landscape. Finally we have experimentally demonstrated that despite the low long range spatial coherence in the multimode state, energy resolved interferometry shows good spatial coherence for the individual modes.

Given the good spatial coherence of the individual modes, it is justified to perform the theoretical analysis in the mean field approximation. We have used a model that describes the polariton dynamics with a generalized Gross-Pitaevskii equation coupled to a rate equation for the density in the exciton reservoir. With this model, we were able to reproduce the main features of the real and reciprocal space densities. The analysis of the condensate modes showed that they are in general different from the linear eigenstates in the disordered potential, but that the lowest energy state is typically very similar to the lowest energy eigenstate within the excitation spot region. The large phase difference of the polariton field between different potential minima explains why the radius of the rings in momentum space correspond to a wavelength of the order of the separation between the potential minima.

We wish to thank Vincenzo Savona for fruitful discussions. This work was supported by EU project STIMSCAT 517769 and EPSRC Grant No. GR/S76076/01 . D. N. Krizhanovskii thanks EPSRC for financial support (EP/E0514448). K. G. Lagoudakis, B. Deveaud-Pl\'{e}dran and M. Richard thank QP-NCCR for financial support through SNSF.


\begin{thebibliography}{23}
\bibitem{FT} Since the precise phase variation across the real space images of the condensed modes is not known, it is not possible to show a precise equivalence between the experimental real and k-space images. See also Section III, where the phase variation in real space is discussed.
\end{thebibliography}
\end{document}